\documentclass[12pt]{article}
\usepackage{amsfonts}
\usepackage{bbm}
\usepackage{graphicx}
\usepackage{booktabs}
\usepackage{subfigure}
\usepackage{mathrsfs}
\usepackage{color}
\usepackage{verbatim}
\usepackage{cancel}
\usepackage{geometry}             
\usepackage{amssymb}
\usepackage{amsmath}
\usepackage{epstopdf}
\usepackage{cases}
\usepackage{cite}
\usepackage{setspace}

\usepackage[hyperindex=true,
pdfstartview=FitH,
bookmarksnumbered=true,
bookmarksopen=true,
citecolor=blue,
linkcolor=blue,
colorlinks=true,
urlcolor=rossoCP3,
unicode]{hyperref}

\definecolor{rossoCP3}{cmyk}{0,.88,.77,.40}

\begin{document}
	\title{\bf \textsf{A shell of Bosons in Spherically Symmetric spacetimes}}
	{\author{\small Duo Li${}^{1}$,\,Bin Wu${}^{1,2,3,4}$\thanks{{\em email}: \href{mailto:binwu@nwu.edu.cn}{binwu@nwu.edu.cn}}{ } , Zhen-Ming Xu${}^{1,2,3,4}$
			, and Wen-Li Yang${}^{1,2,3,4}$
			\vspace{5pt}\\
			\small $^{1}${\it School of Physics, Northwest University, Xi'an 710127, China}\\
			\small $^{2}${\it Institute of Modern Physics, Northwest University, Xi'an 710127, China}\\
			\small $^{3}${\it Shaanxi Key Laboratory for Theoretical Physics Frontiers, Xi'an 710127, China}\\
			\small $^{4}${\it Peng Huanwu Center for Fundamental Theory, Xi'an 710127, China}
		}

\date{}
\maketitle
\begin{spacing}{1.1}
\begin{abstract}
	The thermodynamic properties of a shell of bosons with the inner surface locating at Planck length away from the horizon of Schwarzschild black holes by using statistical mechanics are studied. The covariant partition function of bosons is obtained, from which the Bose-Einstein condensation of bosons is found at a non-zero temperature in the curved spacetimes. As a special case of bosons, we analyze the entropy of photon gas near the horizon of the Schwarzschild black hole, which shows an area dependence similar to the Bekenstein-Hawking entropy. The results may offer new perspectives on the study of black hole thermodynamics. All these are extended to the $D+1$ dimensional spherically symmetric static spacetimes.
\end{abstract}

\section{Introduction}
Bekenstein demonstrated that a black hole has entropy which is proportional to the area of its horizon, starting the study of black hole thermodynamics \cite{Bekenstein 1,Bekenstein 2}. It is the important evidence of black hole thermodynamics that Hawking derived the temperature of a black hole by the method of semi-classical quantum field theory and determined the exact form of the entropy $S=k_Bc^3A/4G\hbar$ \cite{Hawking,Hawking 2}. The black hole thermodynamics is considered as a powerful tool for us to understand the black hole systems and has been an active and fascinating field of research with intriguing results \cite{Carlip:2014pma,Bardeen:1973gs,Wald:1999vt}. 

However, despite the abundant thermodynamic behaviors of various kinds of black holes following the pioneering work of Bekenstein and Hawking, the true nature of black hole thermodynamics is far from being fully understood. The value of black hole entropy is proportional to the area instead of its volume, which is a surprising result and attracts particular attention from physicists. Many efforts have been done to explain the microscopic origin of the black hole entropy. An important development was made in \cite{Srednicki:1993im} by Srednicki, who directly calculated the reduced density and the corresponding entropy (named as entanglement entropy later) in flat spacetime by tracing over the degrees of freedom residing inside an imaginary surface. This method is successfully applied to the case of black hole \cite{Frolov:1993ym}. The entanglement entropy has the intriguing feature that is proportional to the area of the entangling surface.
Later, an efficient approach for the entanglement entropy was developed in Susskind's paper by using replica trick \cite{Susskind:1993ws}. The leading divergence term in the entanglement entropy can be removed by the standard renormalization of Newton’s constant $G$ proposed in \cite{Susskind:1994sm,Jacobson:1994iw}. All divergences are removed in the entropy of the Schwarzschild black hole in  
\cite{Solodukhin:1994yz} by Solodukhin, of the generic static black hole in \cite{Fursaev:1994ea} by Fursaev and Solodukhin (for a nice review, see Ref.\cite{Solodukhin:2011gn}).
It is also suspected that the entanglement entropy can be considered as a quantum contribution to the Bekenstein-Hawking entropy \cite{Das:2008sy}. Besides, several different approaches were proposed to explain the area law of the black hole entropy. In the literature \cite{Strominger:1996sh}, Strominger and Vafa reproduced the Bekenstein-Hawking entropy by counting the number of states of a weakly coupled D-brane system in string theory. The fuzzball theory proposed by Lunin and Mathur is also a speculative candidate to explain the area dependence of Bekenstein-Hawking entropy \cite{Lunin:2002qf}. 

As an inspiring question, it is worth considering that whether the thermodynamic properties (entropy) of black holes are only a superficial coincidence that arises from the microscopic quantum structure of spacetimes or black holes themselves are conventional thermodynamic systems?  To address this question, we must know about the interior information of the black holes, which is a huge obstacle at present. Hence considering the thermodynamic system near the horizon of the black hole seems more significant and would be a helpful approach to give a glimpse of the black hole thermodynamics. In Ref. \cite{Hooft 2}, the ``brick-wall'' model is proposed by 't Hooft, which is a conjecture that the entropy of a black hole comes from the contributions of the quantum gases in a ``wall'' outside the horizon. The ``brick-wall'' model has been greatly generalized to various black holes models \cite{Mann:1990fk,Ghosh:1994wb,Ghosh:1994mm,Ho:1998du,Li}. In fact, it is found that the ``brick-wall'' model is closely related to the entanglement entropy mentioned above \cite{Demers:1995dq}. The UV divergences in the brick-wall entropy are suggested that can be renormalized by the renormalization of the couplings in the gravitational action in the same way as for the entanglement entropy. Besides, the generalized uncertainty principle was considered in \cite{Li 2,Yoon:2007aj,Kim:2007if,Son:2011ii} to replace the cut-off introduced to avoid divergence of the entropy in the brick-wall model.
 
The ``brick-wall'' model is impressive, it tells us the thermodynamic behaviors of the quantum field that obeys the Klein-Gordon equation by using the method of WKB approximation. Naturally, an interesting question is arising as to how a macroscopic thermodynamic system consisting of particles behaves in the background of a black hole. In the literature \cite{Padmanabhan 2}, Padmanabhan considered a box of indistinguishable Boltzmann particles in the background of the spherically symmetric spacetimes and found that the entropy of it is proportional to the logarithm of the transverse area $A$ of the box by using the conventional statistical method. More precisely, the entropy of a box of Boltzmann particles with volume $V$ near the horizon of Schwarzschild black hole behaves like that with volume $AL_p/2$ in the flat spacetime. Such dependence is extended to the stationary axisymmetric spacetimes and cosmological spacetimes \cite{Padmanabhan 3}.
Recently, the area dependence of the entropy of a perfect fluid in a box is also verified via relativistic kinetic theory, provided the bottom of the box is sufficiently close to the Rindler horizon \cite{Hao:2021ifw}.

Moreover, the quantum degrees of freedom on the horizon of black holes are suggested to be described in terms of the non-interacting quasiparticles \cite{Iizuka:2003ad}. The condensation of a two-dimensional box of bosons located on the stretched horizon is also studied and some intriguing results are obtained \cite{Zare}. However, the longitudinal coordinate is automatically omitted in their partition function, and the results are highly restricted to the case that the thermodynamic system should always be two dimensions. In fact, we are naturally interested in a general case that the thermodynamic system is not merely a "plane", but an arbitrary dimensional shell. It is believed that the effect of the longitudinal direction should be taken into account, which could change the thermodynamic behaviors of the bosonic shell. There are two reasons why we choose bosons. i) In light of the rich phenomenology of the well-known Bose-Einstein condensation (BEC), we are interested in the effect of the strong gravitational field on the arbitrary dimensional bosonic system. ii) In line with the Hawking radiation of black holes, in some sense, it can be regarded as the black body radiation which is the behavior of photon gas. The idea is similar to the brick wall model, instead, we consider the photon gas here as non-interacting bosons.

The paper is organized as follows. In section \ref{S2}, we briefly review the calculation of the covariant phase space volume in Schwarzschild spacetime, which is the essential part of the statistic method in curved spacetimes. In section \ref{S3}, we obtain the grand partition function of the bosonic shell near the horizon of the black hole. The critical temperature of the BEC is analyzed, the entropy of photon gas with area dependence is studied. All the results are extended to $D+1$ dimensional spherically symmetric spacetimes (in which the bosonic shell is in $D$ spatial dimensions) in section \ref{S4}. The metric signature is taken as $(-,+,+,+,\cdots)$ and the small letter $a, b$ indices span all spacetime directions, the greek indices $\alpha, \beta$ denote spatial directions. The unit $G=\hbar=c=k_B=1$ is adopted throughout this paper.

\section{Covariant phase space volume}\label{S2}
We consider a shell of bosons surrounding the static spherical black hole to be in thermal equilibrium with the inverse temperature $\beta$ measured by the infinity observers, and we expect that the strong gravitational field would significantly affect the thermodynamic behaviors of the bosonic system. For convenience, it is assumed that neither the interaction between the identical bosons nor the backreaction to the spacetimes exists. The explicit expression of the grand partition function in curved spacetimes plays a crucial role in the analysis of the thermodynamics of bosons, which is the central task in this paper. In this section, we are going to provide a brief introduction to the covariant phase space volume in curved spacetimes at first, and subsequently calculate the grand partition function, from which all the thermodynamic variables can be derived.

The definition of the grand partition function is
\begin{align}
  \ln\Xi=-\int \ln(1-e^{-\beta E-\alpha})D(E)dE,  \label{PF}
\end{align}
where $\beta$ is the inverse temperature, $\alpha=-\beta \mu$, and $\mu$ is the chemical potential.
The distribution of the microscopic states of the bosonic particles is constrained around the energy thin shell in the phase space, the density of states $D(E)$ available to the particles can be obtained from the differential of covariant phase space volume $P(E)$ as \cite{Padmanabhan 1}
\begin{align} 
	D(E)&=\frac{dP(E)}{dE},  \label{DE} \\
	P(E)&=\int_{\Omega} d^3xd^3p, \label{PS}
\end{align}
where the element $d^3xd^3p$ can be proved to be coordinate invariant (the detailed proof is referred to the chapter 7 of literature \cite{Dewitt}). $\Omega$ is the volume of phase space surrounded by the hypersurface described by the mass shell equation $p^ap_a=-m^2$. Now we introduce a timelike killing vector $\xi^a=(1,0,0,0)$ in the static spacetime so that the energy of the bosonic particle is defined by $E=-\xi^a p_a=-p_0$. Expanding the mass shell equation $p^ap_a=-m^2$ leads to $g^{00}{p_{0}}^2+\gamma^{\alpha \beta}p_\alpha p_\beta=-m^2$. Here the spatial metric $\gamma^{\alpha \beta}=g^{\alpha \beta}$ is introduced. We promptly get the relation
\begin{align}\label{E3}
  \gamma^{\alpha \beta}p_\alpha p_\beta=\frac{E^2}{-g_{00}}-m^2.
\end{align}
In order to take the momentum space integration in Eq.(\ref{PS}), we introduce an auxiliary orthogonal matrix $O=\{O_{\mu\nu}\}$ to diagonalize the spatial metric $M=\{\gamma^{\alpha \beta}\}$. 
The left side of Eq.(\ref{E3}) is altered to
\begin{align}
  p^T \,M \, p= p^T \, O^T O M \, O^T O p = \tilde{p}^T D \tilde{p} =\sum_n d_n \tilde{p}_n^2, \nonumber
\end{align}
where the momentum is redefined as $\tilde{p}_\alpha=\sum O_{\alpha \mu}p_\mu$, and  $\{d_n\}$ is the eigenvalue of the diagonalized matrix $OMO^T=D$. The integration element $d^3p$ is equivalently transformed to $d^3\tilde{p}$ since the Jacobian determinant of the orthogonal matrix is identically equal to 1. Further, denoting $p_n'=\sqrt{d_n}\tilde{p}_n$ gives $d^3\tilde{p}=\frac{1}{\sqrt{\det (D)}} d p'^3$ and
\begin{align}\nonumber
  \sum_n (p_n')^2=\frac{E^2}{-g_{00}}-m^2,
\end{align}
which is clearly a 3-sphere in momentum space with radius $R=\sqrt{E^2/(-g_{00})-m^2}$. Thus the phase space volume $P(E)$ arrives at 
\begin{align}
    P(E)&=\int d^3x\frac{1}{\sqrt{\det (D)}}\frac{4\pi}{3}(E^2/(-g_{00})-m^2)^{3/2}\nonumber \\
    &=\int d^3x\sqrt{\gamma}\frac{4\pi}{3}\left(\frac{E^2}{-g_{00}}-m^2\right)^{3/2}, \label{PS2}
\end{align}
where we have used the relation 
\begin{align}\nonumber
  \det (D)=\det(O^TMO)=\det(O^T)\det(M)\det(O)=\det(M)=\gamma^{-1},
\end{align}
and the determinant $\gamma=\rm{det}\{\gamma_{\alpha \beta}\}$.

When the bosonic shell is placed near the horizon of a black hole, we expect its thermodynamic features would be significantly affected by the strong gravitational field, which would promote our understanding of the black hole thermodynamics. The metric function $g_{00}$ tends to zero in the near horizon limit, so that the term $\frac{E^2}{-g_{00}}$ would make the primary contribution to the integration Eq.(\ref{PS2}). We take the series expansion and keep the leading order term, the integration Eq.(\ref{PS2}) approximate to
\begin{align}
P(E) \approx \frac{4\pi}{3}\int d^3 x\sqrt{\gamma}\frac{E^3}{(-g_{00})^{3/2}}.  \label{PSA}
\end{align}
Since the contributions of the mass term become negligible in the near horizon limit, it is notable that phase space volume Eq.(\ref{PSA}) is independent of the mass of the bosons, which is markedly different from the ordinary thermodynamics system in the conventional statistic mechanism.

\section{Bosons in the background of Schwarzschild spacetime}\label{S3}
As a concrete example for us to test the effect of the strong gravitational field on the bosons, the spherically symmetric static spacetime, i.e., the Schwarzschild black hole as the simplest model is considered.
We will first work out the analytical expression of the partition function in the curved spacetime. And then we investigate the existence of the critical temperature for the aggregation of bosons in the background of Schwarzschild black hole, which is an exotic phenomenon known as Bose-Einstein condensation. Subsequently, we study a special bosonic system, the photon gas, whose chemical potential is set to zero, and all the thermodynamic quantities are obtained by using the well-known thermodynamics relations as usual in the conventional statistic mechanism.

The metric of Schwarzschild spacetime is given by
\begin{align}
  ds^2=-\left(1-\frac{2M}{r}\right)dt^2+\left(1-\frac{2M}{r}\right)^{-1}dr^2+r^2(\sin^2\phi d\theta^2+d\phi^2),\nonumber
\end{align}
where $M$ is the mass of black hole, and the position of the horizon is clearly $2M$. The bosonic shell surrounds the back hole with the inner radius $r_a=2M+h$ and outer radius $r_b=2M+H$. The phase space volume Eq.(\ref{PSA}) in the background of Schwarzschild spacetime turns out to be 
\begin{align}
    P(E)&=\frac{4\pi E^3}{3}\int_\Omega \sin\theta d\theta d\phi \int_{r_a}^{r_b}\frac{r^2}{(1-\frac{2M}{r})^2}dr  \nonumber \\
    &=\frac{16\pi^2 E^3}{3}\bigg[\frac{16M^4}{h}-\frac{16M^4}{H}+32M^3\ln\frac{H}{h} \nonumber \\
    &\quad+24M^2(H-h)+4M(H^2-h^2)+\frac{H^3-h^3}{3}\bigg].  \label{PS3}
\end{align}
Since one can not distinguish the inner surface within a Planck length $L_p$ from the horizon \cite{Padmanabhan 4}, a minimal length related to the Planck length $L_p$ is defined as the proper length from the horizon to the inner surface
\begin{align}\nonumber
  L_p=\int_{2M}^{2M+h}\frac{dr}{\sqrt{-g_{00}(r)}}\approx \int_{2M}^{2M+h}\frac{dr}{\sqrt{\frac{r}{2M}-1}} = 2\sqrt{2hM},
\end{align}
where the near horizon approximation has been used $g_{00}(r)\approx (r-2M)g_{00}'(2M)$. Then we get $h=L_p^2/8M$, substituting it into Eq.(\ref{PS3}) yields
\begin{align} 
  P(E)=\frac{16\pi^2 E^3}{3}\left[\frac{128M^5}{L_p^2}-\frac{16M^4}{H}+32M^3\ln\frac{8MH}{L_p^2}+\mathcal{O}(L_p^2)+...\right].  \label{PSS}
\end{align}
Without loss of generality, the thickness of the shell $H-h$, equivalently $H$, is set to satisfy the relation $L_p\ll H\ll 2M$, thus the first term in Eq.(\ref{PSS}) is much larger than others. Omit the higher-order terms and keep the leading order term, the approximation of $P(E)$ becomes
\begin{align}
  P(E)\approx \frac{2\pi E^3A_H}{3L_p^2\kappa^3}, \label{E21}
\end{align}
where $\kappa=1/4M$ is the surface gravity and $A_H=4\pi(2M)^2$ is the area of the horizon. With the help of Eq.(\ref{E21}), the partition function Eq.(\ref{PF}) is given by
\begin{align}
    \ln\Xi&\approx-\frac{2\pi A_H}{3L_p^2\kappa^3}\int_0^\infty \ln(1-e^{-\beta E-\alpha})dE^3    \nonumber \\
    &=\frac{2\pi A_H}{3L_p^2\kappa^3}\int_0^\infty \frac{\beta E^3e^{-\beta E-\alpha}}{1-e^{-\beta E-\alpha}}dE \nonumber  \\
    &=\frac{2\pi A_H}{3L_p^2\kappa^3\beta^3}\int_0^\infty \frac{x^3e^{- x-\alpha}}{1-e^{-x-\alpha}}dx, \label{E11}
\end{align}
where we have used the integration by parts in the second line and denoted $x=\beta E$ for simplification in the third line.
The integral in Eq.(\ref{E11}) can further be expressed as
\begin{align}\nonumber
    \int_0^\infty \frac{x^3e^{- x-\alpha}}{1-e^{-x-\alpha}}dx&=\int_0^\infty \sum_{k=1}^\infty x^3e^{-kx-k\alpha}dx=\sum_{k=1}^\infty \frac{z^k}{k^4}\int_0^\infty y^3e^{-y}dy=6 \sum_{k=1}^\infty \frac{z^k}{k^4},
\end{align}
where $y=kx$, and $0\leq z=e^{-\alpha}\leq 1$ is the fugacity of the bosons. We denote $g_4(z)\equiv\sum_{k=1}^\infty\frac{z^k}{k^4}$ which is the polylogarithm function. The partition function turns out to be
\begin{align}
  \ln\Xi \approx\frac{4\pi A_Hg_4(z)}{L_p^2\kappa^3\beta^3}.\label{E13}
\end{align}
 Before we proceed with the evaluation of thermodynamic quantities from the partition function, we would like to introduce the Tolman law which describes how the temperature in a fixed gravitational field depends on the position. The Tolman temperature is the local temperature measured by the local observers, that is to say, $\beta_{{\rm loc}} (r)=\beta \sqrt{-g_{00}(r)}$ for the inverse temperature \cite{Tolman 1}, where $\beta$ is the inverse temperature measured by the infinity observers. In particular, for the Schwarzschild spacetime, we have
 \begin{align}
  \kappa L_p=\frac{L_p}{4M}\approx\sqrt{\frac{L_p^2}{16M^2+L_p^2}}=\sqrt{1-\frac{2M}{2M+\frac{L_p^2}{8M}}}=\sqrt{-g_{00}(r_a)},\nonumber
 \end{align}
 which transforms the Tolman law to $\beta_{{\rm loc}} (r_a)\approx\kappa L_p\beta$. The partition function Eq.(\ref{E13}) becomes
 \begin{align}
  \ln\Xi\approx \frac{4\pi A_Hg_4(z)}{L_p^2\kappa^3\beta^3}=\frac{4\pi A_HL_p g_4(z)}{(\beta_{{\rm loc}}(r_a))^3}.\label{E15}
 \end{align}
 As mentioned above, once we have the expression of the partition function, all the thermodynamic quantities can be obtained from the well-known thermodynamic relations, that's what we plan to do in the next subsections.  

\subsection{Bose-Einstein Condensation in Schwarzschild spacetime}
Logically, nothing is undetermined once the partition function is given. However, it is worth noting that something is losing during the approximate calculation of the partition function. For example, the particle number is obtained from the partition function Eq.(\ref{E15}) as follows
 \begin{align}
  N= -\frac{\partial}{\partial\alpha} \ln \Xi = z\frac{\partial }{\partial z}\ln \Xi \approx 4\pi A_HL_p T_{\rm loc}(r_a)^3 g_3(z).  \label{N}
 \end{align}
 For a system with the conserved number of particles, the left side of Eq.(\ref{N}) is a constant, and polylogarithm function $g_3(z)$ is a monotone increasing function. With the temperature decreasing, the value of $g_3(z)$ increases. However, there must be a critical temperature $T_c$ when $g_3(z)$  reaches its maximal value with $z=1$. The critical temperature is determined as
\begin{align}\label{E17}
  T_c \approx \left(\frac{N}{4\pi\zeta(3) A_HL_p} \right)^{\frac{1}{3}},
\end{align}
where $g_3(1)=\sum_{k=1}^{\infty} \frac{1}{k^3}=\zeta(3)$ is the Riemann function. Unfortunately, since the limitation on the maximal value of $g_3(z)$ exists, the contradiction emerges when the temperature is lower than the critical temperature and the Eq.(\ref{N}) is no longer satisfied. 

The problem can be solved if we have a careful review of the approximate treatment for the partition function. The original definition of the partition function is $\ln\Xi=-\sum_{s}\omega_s\ln(1-e^{-\beta E - \alpha})$, where the sum is over all the possible states of the system, and $\omega_s$ is the degeneracy of the states. For a macroscopic thermodynamic system, the energy levels are considered as continuous and the partition function is simplified significantly by approximating the summation to an integration. Based on the fact that the energy density $D(E) \sim E^2$ (the energy density is the first derivative of the phase space volume $P(E) \sim E^3$), the contribution from the zero energy state is neglected. It is reasonable for this approximation when the temperature is large. However, the situation changes when the temperature is lower than the critical temperature since the value of the fugacity $z$ is comparable to 1 in this case, as we can see that the contribution from the zero energy state is no longer negligible. For this we need to add an extra term $\ln \Xi_{E=0}=-\ln(1-z)$ to Eq.(\ref{E15}) (where we have set $\omega_0=1$ for convenience)
\begin{align}
  \ln \Xi\approx \frac{4\pi A_HL_p g_4(z)}{(\beta_{\text{loc}}(r_a))^3}-\ln(1-z).\nonumber
\end{align}
Thus the particle number of bosons is
\begin{align}\nonumber
  N&=z\frac{\partial \ln \Xi}{\partial z}=\bar{N_0}+\bar{N}_{\rm exc}, \\
  \bar{N_0}&=\frac{z}{1-z},  \qquad  \bar{N}_{\rm exc}\approx \frac{4\pi A_HL_p g_3(z)}{(\beta_{\text{loc}}(r_a))^3},
\end{align}
where $\bar{N_0}$ is the particle number of the zero energy state and $\bar{N}_{\rm exc}$ is the total particle number of all the excited energy states.
Combined with the formula of the critical temperature Eq.(\ref{E17}), the particle number of the zero energy state is expressed as
\begin{align}
  \bar{N_0}=N-\bar{N}_{\rm exc}=N\left[1-\left(\frac{T_{\rm loc}(r_a)}{T_c}\right)^3\right],      \qquad (T_{\rm loc}(r_a) \leq T_c).
\end{align}
It indicates that with the local temperature decreasing, the total particle number of all the excited states decreases, and the particle number of the zero energy state increase, eventually tend to be $N$ with $T_{\rm loc}(r_a) \rightarrow 0$. This phenomenon is the expected BEC in the curved spacetime. The critical temperature given by the Eq.(\ref{E17}) displays the effect of the strong gravitation field, whose cube is inversely proportional to the area of the horizon. Compared to the results in Ref.\cite{Zare}, in which the square of the critical temperature is inversely proportional to the area of the horizon, the power law of the critical temperature is changed attributing to the effect of the longitudinal direction of the thermodynamic system.

If we reconsider the Tolman law ( $\beta_{\rm loc}(r_a)\approx\kappa L_p\beta$ and $T_0=1/\beta$), i.e., the critical Tolman temperature is related to the temperature measured by the infinity observers by $T_0\approx \kappa L_pT_c$, we have  
\begin{align}
  T_0\approx T_H\left(\frac{2\pi^2L_p^2N}{\zeta(3)A_H}\right)^{\frac{1}{3}},
\end{align}
where the formula of the Hawking temperature $T_H=\kappa/2\pi$ is used. It suggests that for the infinity observers, the critical temperature of BEC is connected with Hawking temperature by a coefficient. The interesting result is obtained if the particle number of bosons is of the same order of the microscopic degrees of freedom of the black hole proposed in the literature by Ruppeiner \cite{Ruppeiner2}, that is to say, if
\begin{align}
  N =\frac{\zeta(3)A_H}{2\pi^2L_p^2}  \sim A_H/L_p^2,\nonumber
\end{align}
the critical temperature of BEC will be equal to the Hawking temperature of the black hole, which indicates that the condensation of the bosonic shell near the horizon would be a useful toy model and might cast new insight into the underlying microscopic structure of the spacetimes.

Next, the entropy of the bosonic shell near the horizon can be directly worked out, which shows
\begin{align}\nonumber
    S= \left(1-\beta_{\rm loc} \frac{\partial}{\partial\beta_{\rm loc}}\right)\ln\Xi\approx\frac{16\pi L_pg_4(z)}{(\beta_{\rm loc}(r_a))^3}A_H.
\end{align}
It suggests that as the bosonic shell with the inner radius locating at Plank length $L_p$ away from the horizon, the entropy
behaves as though the shell possesses the volume $L_p A_H$ rather than the total volume $V$. The reason is that the contribution to entropy mainly comes from the degrees of freedom very close to the horizon and others away from the horizon are relatively negligible. The entropy shows an area of the system dependence instead of the volume dependence as the standard property known in ordinary thermodynamic systems. In the next section, we are going to investigate the situation of $\mu=0 \,(z=1)$, that is photon gas, which may have profound meanings.

\subsection{Entropy of photon gas}
When particular attention has been paid to photon gas, whose static mass vanishes, the mass term automatically disappears in the phase space volume, and all the calculation of the partition function above is also valid. Since the chemical potential of the photon gas is always be zero, the polylogarithm function $g_4(z)$ becomes the Riemann function $\zeta (4)=\sum_{k=1}^\infty\frac{1}{k^4}=\frac{\pi^4}{90}$, and the partition function arrives at
\begin{align}
  \ln\Xi\approx \frac{4\pi A_HL_p}{(\beta_{{\rm loc}}(r_a))^3}\frac{\pi^4}{90}=\frac{2\pi^5A_HL_p}{45(\beta_{{\rm loc}}(r_a))^3}.\nonumber
\end{align}
The internal energy and the entropy of the photon gas read as
\begin{align}
   U&= -\frac{\partial}{\partial\beta_{\rm loc}}\ln\Xi\approx\frac{2\pi^5A_HL_p}{15(\beta_{{\rm loc}}(r_a))^4},
    \\
  S&= \left(1-\beta_{\rm loc} \frac{\partial}{\partial\beta_{\rm loc}}\right)\ln\Xi\approx \frac{8\pi^5A_HL_p}{45(\beta_{\rm loc}(r_a))^3}. 
  \label{E23}
\end{align}
As we can see that the internal energy is proportional to the bi-quadratic of the temperature, which is consistent with the Stefan-Boltzmann law. The proportional coefficient, i.e., the Stefan-Boltzmann constant can be exactly read off from the above equation, which is related to the area of the black hole horizon. A little further, if we assume that the photon shell is in equilibrium with its surroundings, the inverse temperature of the thermal bath $\beta$ measured by the infinity observers can be valued as the Hawking inverse temperature $\beta=2\pi/\kappa $. So that the inverse Tolman temperature measured by the observers located at $r_a$ is approximate to $\beta_{\rm loc}(r_a)\approx2\pi L_p$, which alters the entropy Eq.(\ref{E23}) result in
\begin{align}
  S\approx \frac{\pi^2A_H}{45(L_p)^2}.\nonumber
\end{align}
We can see that in Schwarzschild spacetime, the entropy of photon gas shows an area dependence with the near horizon limit. Furthermore, with the help of the relation $L_p=\sqrt{G\hbar/c^3}$, we finally get the entropy of photon gas as the form
\begin{align}
  S\approx \frac{\pi^2c^3A_H}{45G\hbar}\to \frac{A_Hc^3}{4G\hbar}.
\end{align}
Coincidentally, we find that the entropy of the photon gas  near the horizon (the leading order term) behaves like the entropy of the
black hole. This feature may be somehow enlightening and offer new perspectives on the study of thermodynamic and quantum properties of black holes.

\section{Bosons in the background of spherically symmetric spacetimes}\label{S4}
In this section, we would like to generalize our results into arbitrary dimensional spacetimes. The metric of $D+1$ dimensional spherically symmetric static spacetimes is
\begin{align}
  ds^2=-f(r)dt^2+f(r)^{-1}dr^2+r^{D-1}d\Omega^2_{D-1},\nonumber
\end{align}
where $d\Omega^2_{D-1}$ is the line element of $(D-1)$-dimensional unit sphere. The phase spacetime volume with the near horizon limit in $D+1$ dimensional spherically symmetric spacetimes is given as \cite{Padmanabhan 2}
\begin{align}
  P(E)\approx\frac{\pi^{D/2}E^D}{\Gamma (\frac{D}{2}+1)}\frac{A_{D-1}}{(D-1)L_p^{D-1}\kappa^D},
\end{align}
where $A_{D-1}$ is the area of the horizon, $\kappa$ is the surface gravity of $D+1$-dimensional black hole.

The logarithm of the partition function $\ln \Xi$ is then
\begin{align}
   \nonumber \ln \Xi&\approx-\frac{\pi^{D/2}}{\Gamma (\frac{D}{2}+1)}\frac{A_{D-1}}{(D-1)L_p^{D-1}\kappa^D}\int \ln (1-ze^{-\beta E})dE^D \\
    &=\frac{\pi^{D/2}D!}{\Gamma (\frac{D}{2}+1)}\frac{A_{D-1}L_p}{(D-1)(\beta_{\rm loc}(r_a))^D}g_{D+1}(z).
\end{align}
Similar to the situation in Schwarzschild spacetime, the Tolman law is used to introduce the local temperature $\beta_{\rm loc}(r_a)\approx\beta \kappa L_p$. The arbitrary dimensional polylogarithm function is defined as $g_{D+1}(z)=\sum_{i=1}^\infty\frac{z^i}{i^{D+1}}$. To analyze the BEC, we should recover the contribution from the zero-energy state to the partition function, that is
\begin{align}
  \ln \Xi\approx\frac{\pi^{D/2}D!}{\Gamma (\frac{D}{2}+1)}\frac{A_{D-1}L_p}{(D-1)(\beta_{\rm loc}(r_a))^D}g_{D+1}(z)-\ln(1-z).\nonumber
\end{align}
The number of bosons is given by
\begin{align}
  N&=z\frac{\partial \ln \Xi}{\partial z}=\bar{N}_{0}+\bar{N}_{\rm exc}, \nonumber \\
  \bar{N}_{0}&=\frac{z}{1-z},  \qquad \bar{N}_{\rm exc}\approx\frac{\pi^{D/2}D!}{\Gamma (\frac{D}{2}+1)}\frac{A_{D-1}L_p}{(D-1)(\beta_{\rm loc}(r_a))^D}g_{D}(z)\nonumber.
\end{align}

The critical temperature (local) $T_c$ is obtained by making $\bar{N}_{\rm exc}=N$ and $z=1$, which leads to
\begin{align}
  T_c\approx\left(\frac{\Gamma (\frac{D}{2}+1)(D-1)}{\pi^{D/2}D!\zeta(D)}\frac{N}{A_{D-1}L_p}\right)^{\frac{1}{D}},
\end{align}
where $\zeta(D)$ is the Riemann function replacing the polylogarithm function. Here we can also see that the power law of the critical temperature with respect to the area of the horizon is related to the dimensions of the thermodynamic system. And we can find the relationship between the critical temperature (red shit to infinity) and Hawking temperature as
\begin{align}
  T_0\approx T_H\left(\frac{2^D\pi^{D/2}\Gamma (\frac{D}{2}+1)(D-1)}{D!\zeta(D)}\frac{NL_p^{D-1}}{A_{D-1}}\right)^{\frac{1}{D}}.
\end{align}
The notable thing is that the critical temperature of BEC is equal to the Hawking temperature when the number of the bosons is set to be
\begin{align}
  N=\frac{D!\zeta(D)}{2^D\pi^{D/2}\Gamma (\frac{D}{2}+1)(D-1)}\frac{A_{D-1}}{L_p^{D-1}},
\end{align}
where $A_{D-1}/{L_p^{D-1}}$ is assumed to be the degrees of freedom of the $D+1$-dimensional black hole.

What's more, when the spatial dimension of the spacetime is $D=2$, we get
\begin{align*}
  T_0 \approx T_H\left(\frac{2\pi N L_p}{\zeta(2) A}\right)^{1/2}.
\end{align*}
The intriguing result shows that for a spatial two-dimensional bosonic system, the critical temperature is nonzero, which indicates that BEC exists. As we know that the BEC for the free bosons in a two-dimensional plane at the nonzero temperature is absent, while the strong gravitational field would affect the critical temperature and make BEC possible at the nonzero temperature.

If we turn the research object to photon gas $(z=1)$, the value of polylogarithm function $g_{D+1}(z)$ becomes a Riemann function $\zeta(D+1)$. So the partition function is simplified to
\begin{align}
  \ln\Xi\approx \frac{\pi^{D/2}D!}{\Gamma (\frac{D}{2}+1)}\frac{A_{D-1}L_p}{(D-1)(\beta_{\rm loc}(r_a))^D} \zeta (D+1).\nonumber
\end{align}
Thus the entropy of photon gas in the $D+1$ dimensional spherically symmetric spacetimes is
\begin{align}
  S\approx \frac{\pi^{D/2}(D+1)!}{\Gamma (\frac{D}{2}+1)}\frac{A_{D-1}L_p}{(D-1)(\beta_{\rm loc}(r_a))^D} \zeta (D+1).
\end{align}
When the photon gas come to thermal equilibrium at Hawking temperature $\beta=2\pi/\kappa $, the expression of entropy reduces to
\begin{align}
  S\approx \frac{(D+1)!}{\Gamma (\frac{D}{2}+1)2^D\pi^{D/2}}\frac{A_{D-1}}{(D-1)L_p^{D-1}} \zeta (D+1).
\end{align}
It is the expression of the entropy of the photon shell in $D+1$ dimensional spherically symmetric static spacetimes. The discussion at the end of the previous section is also applicable here, the entropy shows an area dependence in $D+1$ dimensions, which provides some insights into the thermodynamics of black holes.

\section{Summary and discussion}
In this paper, we studied the thermodynamic behaviors of a shell of bosons near the horizon of a spherically symmetric black hole based on the method of statistical mechanics. We find that the bosons begin to condense near the horizon of Schwarzschild spacetime from a non-zero Tolman temperature $T_c$ in Eq.(\ref{E17}), which is known as Bose-Einstein condensation. In particular, the critical temperature is connected with the Hawking temperature by a coefficient for the observers located at infinity. Furthermore, the critical temperature of BEC will be equal to the Hawking temperature of the black hole when the particle number of bosons is of the same order as the microscopic degrees of freedom of the black hole, which implies that the condensation of the bosonic shell near the horizon would be a useful toy model to shed new lights on the underlying microscopic structure of the spacetimes.

In addition, we obtained the entropy of the bosonic shell from its grand partition function, which shows an area dependence instead of volume dependence. When the fugacity $z$ is equal to $1$, that is to say, the object we focusing on becomes photon gas, the entropy derived from the same procedure behaves like the Bekenstein-Hawking entropy, which may bring some new perspectives for the investigations of the black hole thermodynamics and quantum gravity. The results and discussions above still hold when we change the background spacetime into $D+1$ dimensional spherically symmetric spacetimes. Surprisingly, the BEC at a non-zero temperature also exists near the horizon in $2+1$ dimensional spacetimes, while the condensation is absent in the two-dimensional plane unless the temperature is zero.

There is seemly no obstacle for us to investigate the thermodynamic behaviors of fermions in the curved spacetime if the Fermi-Dirac statistic is considered. It is worth expecting how the influence of the strong gravitational field would be on the degenerate electron gas and the entropy of fermions. We look forward to more enlightening and intriguing results and leave this for future works.

\section*{Acknowledgment}
We would like to thank Wei Xu and Chao Wang for their useful discussions. The financial supports from the National Natural Science Foundation of China (Grant Nos.12047502, 11947208), the China Postdoctoral Science Foundation (Grant Nos.2017M623219, 2020M673460), Basic Research Program of Natural Science of Shaanxi Province (Grant No.2019JQ-081) and Double First-Class University Construction Project of Northwest University are
gratefully acknowledged.

\end{spacing}
\begin{spacing}{1}

\end{spacing}
\end{document}